\documentclass[ preprint,   aip,cha,]{revtex4-1}
\usepackage{amssymb}
\usepackage{amsmath}
\usepackage{docs}
\usepackage{bm}
\usepackage[colorlinks=true,linkcolor=blue]{hyperref}
\usepackage{graphicx}

\expandafter\ifx\csname package@font\endcsname\relax\else
 \expandafter\expandafter
 \expandafter
\expandafter
 \expandafter{\csname package@font\endcsname}
\fi
\input{tcilatex}

\begin{document}

\title{Intrinsic fluctuations of chemical reactions with different approaches}%

\author{Hong-Yuan Xu}%
\affiliation{Department of Physics, Chung-Yuan Christian University, Chungli, 32023 Taiwan}%
\author{Yu-Pin Luo}%
\affiliation{Department of Electronic Engineering, National Formosa University, Huwei, 63201 Taiwan}%
\author{Ming-Chang Huang}%
\email{mchuang@cycu.edu.tw}
\affiliation{Department of Physics, Chung-Yuan Christian University, Chungli, 32023 Taiwan}%

\begin{abstract}
The Brusselator model are used for the study of the intrinsic fluctuations
of chemical reactions with different approaches. The equilibrium states of
systems are assumed to be spirally stable in mean-field description, and two
statistical measures of intrinsic fluctuations are analyzed by different
theoretical methods, namely, the master, the Langevin, and the linearized
Langevin equation. For systems far away from the Hopf bifurcation line, the
discrepancies between the results of different methods are insignificant
even for small system size. However, the discrepancies become noticeable
even for large system size when systems are closed to the bifurcation line.
In particular, the statistical measures possess singular structures for
linearized Langevin equation at the bifurcation line, and the singularities
are absent from the simulation results of the master and the Langevin
equation.
\end{abstract}
%\date{March 2010}%
%\revised{August 2010}%

\maketitle

%\tableofcontents

\section{Introduction}

\label{s0b}

The mean-field descriptions of molecular reactions have been very effective
in studying the macroscopic features of a chemical system. In general, a
mean-field model is described by a set of rate equations, and the stability
of equilibrium state may vary with adjustable parameters in the model. Hopf
bifurcation is a type of bifurcations for which, the stability of
equilibrium state switches and a periodic solution arises as a small smooth
variation in the values of parameters is made \cite{crawford,arnold}. A
concrete example can be given by the Brusselator model of chemical
reactions, which is a theoretical model demonstrating the existence of the
phase of oscillating reactions \cite{strogatz,tomita,boland}. However, an
important factor is absent from the mean-field consideration, namely, the
stochasticity in chemical reactions, it arises because of the finite number
of molecules and the probabilistic feature of reactions. The stochasticity
may be smoothed out for systems with large number of molecules, but it is
definitely important for small systems \cite%
{mckane,moreira,lama,scott,moss,kampen,sagues}.

Many theoretical methods have been developed to analyze the stochastic
effect and the related problems in chemical reactions. The probability
density distribution of molecular numbers can be studied by means of the
chemical master equation at the level of individual molecules. The chemical
master equation is discrete, and one of the approximated continuous
equations is the Fokker-Planck equation, obtained from the Kramers-Moyal
expansion of the chemical master equation \cite{kampen,gardiner,risken}.
Alternatively, equivalent to Fokker-Planck equation one can also use
stochastic differential equation, called Langevin equation, to study the
stochastic effect \cite{gardiner,risken}; moreover, the equation is often
linearized about the equilibrium state of mean-field equation for analytic
study. Along with this main stream, novel methods have been developed for
specific studies \cite{gaspard,nakanishi,broeck,vance,mori}. For example,
based on the chemical master equation Gaspard used the Hamilton-Jacobi
method to give a formalism for the study of oscillating reactions \cite%
{gaspard}, and Nakanishi and et al. employed the formalism to analyze the
molecular density distribution in a chemical oscillator \cite{nakanishi}.
Among different approaches, it is essential to understand the adequacy and
the limitation of a method. An example can be given by a recent study in
microbial biology: The stable coexistence state of the deterministic kill
the winner model can be destroyed by demographic stochasticity, however, the
diversity of the ecosystem can be maintained in a stochastic model of the
coevolution at the level of individual species \cite{xue}. This motivates us
to look into the discrepancy in the statistic measures of intrinsic
fluctuations between different theoretical approaches.

We take the Brusselator model as the working frame for chemical reactions in
this work. In the model, the Hopf bifurcation line separates the mean-field
equilibrium states into two types, spirally stable states and spirally
unstable states \cite{strogatz,tomita,boland}. In this work, we focus on the
spirally stable states and investigate two statistical measures of intrinsic
fluctuations, steady-state probability density distributions and power
spectra, with three different approaches, the master, the Langevin, and the
linearized Langevin equation. The discrepancy between the results is
analyzed by considering two factors, the system size and the distance of
equilibrium state from the bifurcation line. The latter is shown to play an
important role in determining the adequacy of a method, in particular, the
analytic results obtained from linearized Langevin equations possess
singular structures at the bifurcation line. \

This paper is organized as follows: In Sec. 2, we first introduce the
Brusselator model and set up the corresponding formulations, the master, the
Fokker-Planck, and the Langevin equations. Among the formulations, two
different Langevin equations correspond to the same Fokker-Planck equation.
In Sec. 3, both Langevin equations are linearized about the equilibrium
state, and the linearized equations lead to the same analytic expressions
for two statistical measures of intrinsic fluctuations. In Sec. 4, we report
the numerical results of the statistical measures obtained from the master,
the Langevin, and the linearized Langevin equations, and the discrepancy
between the results is discussed. Finally, we summarize the obtained results
in Sec. 5. \ \

\label{s0e}\ \

\section{Formulations of Brusselator Model \ \ }

\label{s1b}

The Brusselator model at the level of individual molecules is defined by
four chemical reactions between four types of reactants, denoted as $A$, $B$%
, $X_{1}$, and $X_{2}$. However, the model was designed in a way that only
the numbers of $X_{1}$ and $X_{2}$ reactants vary with time, meanwhile the
numbers of $A$ and $B$ maintain constant to set the reaction rates \cite%
{strogatz,tomita,boland}. The reactions are
\begin{eqnarray}
R_{1} &:&\text{ }A\rightarrow X_{1}+A,\text{ }  \label{001} \\
R_{2} &:&\text{ }X_{1}\rightarrow \emptyset ,\text{ }  \label{002} \\
R_{3} &:&\text{ }X_{1}+B\rightarrow X_{2}+B,  \label{003} \\
R_{4} &:&\text{ }2X_{1}+X_{2}\rightarrow 3X_{1};  \label{004}
\end{eqnarray}%
and the state of the system at time $t$ is described by respective number of
$X_{1}$ and $X_{2}$ molecules at the moment, denoted as $\mathbf{n}^{\tau
}\left( t\right) =\left( n_{1}\left( t\right) ,n_{2}\left( t\right) \right) $%
. Note that the boldfaced letters, hereafter, are used to indicate matrices
with the superscript $\tau $ for the transpose. When the reactions occur the
state $\mathbf{n}$ will change; the vector $\mathbf{u}_{\left( j\right) }$
is introduced to specify the change of molecular numbers caused by the
occurrence of a $R_{j}$ reaction. By observing the reactions given by Eqs. (%
\ref{001}) - (\ref{004}), we have
\begin{equation}
\mathbf{u}_{\left( 1\right) }^{\tau }=\left( 1,0\right) ,\text{ }\mathbf{u}%
_{\left( 2\right) }^{\tau }=\left( -1,0\right) ,\text{ }\mathbf{u}_{\left(
3\right) }^{\tau }=\left( -1,1\right) ,\text{ and }\mathbf{u}_{\left(
4\right) }^{\tau }=\left( 1,-1\right) .  \label{005}
\end{equation}

We further specify the transition rate of a channel to give a complete
characterization of the reactions. The transition rate of $R_{j}$ channel,
denoted as $\Gamma _{j}\left( \mathbf{n}\right) $ for $j=1,\cdot \cdot \cdot
,4$, takes the mathematical form, $\Gamma _{j}\left( \mathbf{n}\right)
=k_{j}h_{j}\left( \mathbf{n}\right) $, where the factor $k_{j}$ is given as
the probability per unit time for a randomly chosen pair of $R_{j}$
reactants to react accordingly, and the factor $h_{j}\left( \mathbf{n}%
\right) $ is the number of combinatory ways between the $R_{j}$ reactants
available in the state $\mathbf{n}$. We follow Ref. \cite{boland} to set up
the transition rates as follows. The $R_{1}$ reaction corresponds to the
spontaneous creation of $X_{1}$ molecules. By parameterizing the number of $%
A $ molecules as the integer $N$, we have $\Gamma _{1}\left( \mathbf{n}%
\right) =N$. The $R_{2}$ reaction signifies the decay of $X_{1}$ molecules,
and it can be used to set the time scale of the model, Then, the transition
rate of the spontaneous decay $R_{2}$ is given as $\Gamma _{2}\left( \mathbf{%
n}\right) =n_{1}$. The $R_{3}$ reaction converts the molecules of $X_{1}$
type to that of $X_{2}$ type. Based on Eq. (\ref{005}) we set the transition
rate as $\Gamma _{3}\left( \mathbf{n}\right) =bn_{1}$, where the parameter $%
b $ contains a factor given by the ratio of the number of $B$ molecules to
that of $A$ molecules. Finally, the $R_{4}$ reaction converts a $X_{2}$
molecule to $X_{1}$ with the transition rate given as $\Gamma _{4}\left(
\mathbf{n}\right) =cN^{-2}n_{1}^{2}n_{2}$ for which, we use $n_{1}^{2}$ to
approximate $n_{1}\left( n_{1}-1\right) $ for $n_{1}\gg 1$ and the factor $%
N^{-2}$ is added to make $b$ and $c$ to have the same dimension.

\subsection{Master and Fokker-Planck equations}

The dynamics of the model can be described by the master equation in which,
the time-evolution of the condition probability, $P\left( \left. \mathbf{n}%
_{t},t\right\vert \mathbf{n}_{0}\mathbf{,}t_{0}\right) $ defined as the
probability for $\mathbf{n}(t)=\mathbf{n}_{t}$ given $\mathbf{n}(t_{0})=%
\mathbf{n}_{t_{0}}$, is given as
\begin{eqnarray}
\frac{\partial }{\partial t}P\left( \left. \mathbf{n}_{t},t\right\vert
\mathbf{n}_{0}\mathbf{,}t_{0}\right) &=&\sum_{j=1}^{4}\Gamma _{j}\left(
\mathbf{n}_{t}\mathbf{-u_{\left( j\right) }}\right) P\left( \left. \mathbf{n}%
_{t}\mathbf{-u_{\left( j\right) }},t\right\vert \mathbf{n}_{0}\mathbf{,}%
t_{0}\right)  \notag \\
&&-\sum_{j=1}^{4}\Gamma _{j}\left( \mathbf{n}_{t}\right) P\left( \left.
\mathbf{n}_{t},t\right\vert \mathbf{n}_{0}\mathbf{,}t_{0}\right) .
\label{007}
\end{eqnarray}

In general, the master equation is hard to manage, and approximations are
often made for analytic study. By observing that the components of $\mathbf{n%
}_{t}$ are very large compared to $1$, we can use the Taylor's expansion to
write
\begin{eqnarray}
f_{j}\left( \mathbf{n}_{t}\mathbf{-u_{\left( j\right) }}\right)
&=&f_{j}\left( \mathbf{n}_{t}\right) +\sum_{i=1}^{2}\left( -u_{\left(
j\right) ,i}\right) \frac{\partial f_{j}\left( \mathbf{n}_{t}\right) }{%
\partial n_{t,i}}  \notag \\
&&+\frac{1}{2}\sum_{i,k=1}^{2}\left( -u_{\left( j\right) ,i}\right) \left(
-u_{\left( j\right) ,k}\right) \frac{\partial ^{2}f_{j}\left( \mathbf{n}%
_{t}\right) }{\partial n_{t,i}\partial n_{t,k}}+\cdot \cdot \cdot ,
\label{008}
\end{eqnarray}%
where $u_{\left( j\right) ,i}$ is the $i$th component of the change vector $%
\mathbf{u}_{\left( j\right) }$, and $n_{t,i}$ is the $i$th component of the
state vector at time $t$, $\mathbf{n}_{t}$. As the expansion is applied to
the master equation of Eq. (\ref{007}), we have the Kramers--Moyal equation
\cite{kampen,risken}. By keeping up to the order of $\left( u_{\left(
j\right) ,k}\right) ^{2}$ and neglecting the higher order terms in
Kramers-Moyal equation, we can obtain the Fokker-Planck equation \cite%
{risken}. Note that the integer $N$, the number of $A$ molecules which is
constant in time, in fact, control the number of molecules in the system,
and we can effectively treat $N$ as the system size. Then, we use the
"molecular concentrations", $\mathbf{x}^{\tau }=\mathbf{n}^{\tau }/N=\left(
x_{1},x_{2}\right) $ with $x_{1}=n_{1}/N$ and $x_{2}=n_{2}/N$, as variables
to the Fokker-Planck equation as
\begin{eqnarray}
\frac{\partial }{\partial t}P\left( \left. N\mathbf{x}_{t},t\right\vert N%
\mathbf{x}_{0}\mathbf{,}t_{0}\right) &=&-\sum_{i=1}^{2}\frac{\partial }{%
\partial x_{t,i}}\left[ \mu _{i}\left( \mathbf{x}_{t}\right) P\left( \left. N%
\mathbf{x}_{t},t\right\vert N\mathbf{x}_{0}\mathbf{,}t_{0}\right) \right]
\notag \\
&&+\sum_{i,k=1}^{2}\frac{\partial ^{2}}{\partial x_{t,i}\partial x_{t,k}}%
\left[ D_{i,k}\left( \mathbf{x}_{t}\right) P\left( \left. N\mathbf{x}%
_{t},t\right\vert N\mathbf{x}_{0}\mathbf{,}t_{0}\right) \right] ,
\label{013}
\end{eqnarray}%
where $\mathbf{\mu }\left( \mathbf{x}_{t}\right) $ is the drift vector
defined as
\begin{equation}
\mathbf{\mu }\left( \mathbf{x}_{t}\right) =\left(
\begin{array}{c}
1-x_{t,1}-bx_{t,1}+cx_{t,1}^{2}x_{t,2} \\
bx_{t,1}-cx_{t,1}^{2}x_{t,2}%
\end{array}%
\right) ,  \label{014}
\end{equation}%
and $\mathbf{D}\left( \mathbf{x}_{t}\right) $ is the diffusion matrix given
as
\begin{equation}
\mathbf{D}\left( \mathbf{x}_{t}\right) =\left( \frac{1}{2N}\right) \left(
\begin{array}{cc}
1+x_{t,1}+bx_{t,1}+cx_{t,1}^{2}x_{t,2} & -bx_{t,1}-cx_{t,1}^{2}x_{t,2} \\
-bx_{t,1}-cx_{t,1}^{2}x_{t,2} & bx_{t,1}+cx_{t,1}^{2}x_{t,2}%
\end{array}%
\right) .  \label{015}
\end{equation}

\subsection{Master to Langevin equation}

One can set up the Langevin equation from the master equation of Eq. (\ref%
{007}). A general construction frame was given explicitly by Gillespie \cite%
{gillespie1,gillespie2}. Here, we follow the frame given by Ref. \cite%
{gillespie1} to construct the Langevin equation as follows. Based on Eq. (%
\ref{007}) we can write
\begin{equation}
n_{i}\left( t+\tau \right) =n_{t,i}+\sum_{j=1}^{4}u_{\left( j\right)
,i}K_{j}\left( \mathbf{n}_{t},\tau \right) ,\text{ }i=1,2,  \label{023a}
\end{equation}%
where $\mathbf{n}_{t}$ and $\mathbf{n}\left( t+\tau \right) $ are the state
of the system at the current time $t$ and the subsequent time $t+\tau $, and
$K_{j}\left( \mathbf{n}_{t},\tau \right) $ denotes the number of $R_{j}$
reactions occurring in the time interval $\left[ t,t+\tau \right] $. For
obtaining an explicit expression of $K_{j}\left( \mathbf{n}_{t},\tau \right)
$, we first assume that the time interval $\tau $ is small enough that the
transition rate $\Gamma _{j}\left( \mathbf{n}_{s}\right) $ for any $s\in %
\left[ t,t+\tau \right] $ can be approximated by $\Gamma _{j}\left( \mathbf{n%
}_{t}\right) $. Then, the events of reactions in the time interval $\left[
t,t+\tau \right] $ are independent of each other, and the numbers of events
for different reaction channels, $K_{j}\left( \mathbf{n}_{t},\tau \right) $,
become statistically independent Poissonian random variables for which, we
denote as $\mathit{P}\left( \Gamma _{j}\left( \mathbf{n}_{t}\right) ,\tau
\right) $ for the $j$th channel. Then, Eq. (\ref{023a}) becomes
\begin{equation}
n_{i}\left( t+\tau \right) =n_{t,i}+\sum_{j=1}^{4}u_{\left( j\right) ,i}%
\mathit{P}\left( \Gamma _{j}\left( \mathbf{n}_{t}\right) ,\tau \right) ,%
\text{ }i=1,2.  \label{024a}
\end{equation}%
Note that the Poissonian random variable $\mathit{P}\left( \Gamma _{j}\left(
\mathbf{n}_{t}\right) ,\tau \right) $ is the number of $R_{j}$ reaction in
the time interval $\left[ t,t+\tau \right] $ with the probability of
occurring a $R_{j}$ reaction in infinitesimal time interval $\left[
t,t+d\tau \right] $ given by $\Gamma _{j}\left( \mathbf{n}_{t}\right) d\tau $%
.

It was shown that the probability for $\mathit{P}\left( \Gamma _{j}\left(
\mathbf{n}_{t}\right) ,\tau \right) $ taking the integer value $n$, denoted
as $\mathit{Q}(n;\Gamma _{j}\left( \mathbf{n}_{t}\right) ,\tau )$, possesses
the form,
\begin{equation}
\mathit{Q}(n;\Gamma _{j}\left( \mathbf{n}_{t}\right) ,\tau )=\frac{\left[
\Gamma _{j}\left( \mathbf{n}_{t}\right) \tau \right] ^{n}\exp \left( -\Gamma
_{j}\left( \mathbf{n}_{t}\right) \tau \right) }{n!},\text{ }n=0,1,2,\cdot
\cdot \cdot ,  \label{025a}
\end{equation}%
and this yields the mean and the variance of $\mathit{P}\left( \Gamma
_{j}\left( \mathbf{n}_{t}\right) ,\tau \right) $ as the same value, $\Gamma
_{j}\left( \mathbf{n}_{t}\right) \tau $, that is,$\qquad $
\begin{equation}
\left\langle \mathit{P}\left( \Gamma _{j}\left( \mathbf{n}_{t}\right) ,\tau
\right) \right\rangle =\sigma ^{2}\left( \mathit{P}\left( \Gamma _{j}\left(
\mathbf{n}_{t}\right) ,\tau \right) \right) =\Gamma _{j}\left( \mathbf{n}%
_{t}\right) \tau .  \label{026a}
\end{equation}%
The probability $\mathit{Q}(n;\Gamma _{j}\left( \mathbf{n}_{t}\right) ,\tau
) $ of Eq. (\ref{025a}) can be further approximated as
\begin{equation}
\mathit{Q}(n;\Gamma _{j}\left( \mathbf{n}_{t}\right) ,\tau )\simeq \left(
2\pi \Gamma _{j}\left( \mathbf{n}_{t}\right) \tau \right) ^{-1/2}\exp -\frac{%
\left( n-\Gamma _{j}\left( \mathbf{n}_{t}\right) \tau \right) ^{2}}{2\Gamma
_{j}\left( \mathbf{n}_{t}\right) \tau },  \label{027a}
\end{equation}%
if we impose an additional condition, namely, although the time interval $%
\tau $ is small but it is large enough to hold the inequality, $\Gamma
_{j}\left( \mathbf{n}_{t}\right) \tau \gg 1$ for $j=1,\cdot \cdot \cdot ,4$.
The form of $\mathit{Q}(n;\Gamma _{j}\left( \mathbf{n}_{t}\right) ,\tau )$
given by Eq. (\ref{027a}) allows us to rewrite Eq. (\ref{024a}) as
\begin{equation}
n_{i}\left( t+\tau \right) =n_{t,i}+\sum_{j=1}^{4}u_{\left( j\right) ,i}%
\mathit{N}\left( \Gamma _{j}\left( \mathbf{n}_{t}\right) \tau ,\Gamma
_{j}\left( \mathbf{n}_{t}\right) \tau \right) ,\text{ }i=1,2,  \label{028a}
\end{equation}%
where $\mathit{N}(m,\sigma ^{2})$ is the normal random variable with mean $m$
and variance $\sigma ^{2}$. Moreover, based on the linear combination
theorem, we have the equality,
\begin{equation}
\mathit{N}(m,\sigma ^{2})=m+\sigma \mathit{N}(0,1).  \label{029a}
\end{equation}%
Consequently, Eq. (\ref{028a}) becomes \cite{gillespie3}
\begin{equation}
n_{i}\left( t+\tau \right) =n_{t,i}+\sum_{j=1}^{4}u_{\left( j\right)
,i}\Gamma _{j}\left( \mathbf{n}_{t}\right) \tau +\sum_{j=1}^{M}u_{\left(
j\right) ,i}\sqrt{\Gamma _{j}\left( \mathbf{n}_{t}\right) }\mathit{N}\left(
0,1\right) \tau ^{1/2}.  \label{030a}
\end{equation}%
Note that the two imposed conditions on the time interval $\tau $, one leads
to Eq. (\ref{024a}) and the other leads to Eq. (\ref{027a}), require $\tau $
to be macroscopic infinitesimal.

The result of Eq. (\ref{030a}) implies the Langevin equation, in terms of
"molecular concentrations", as
\begin{equation}
\frac{dx_{i}\left( t\right) }{dt}=\mu _{i}\left( \mathbf{x}\right) +\frac{1}{%
\sqrt{N}}\sum_{j=1}^{4}A_{ij}\left( \mathbf{x}\right) \zeta _{j}\left(
t\right) ,\text{ }i=1,2  \label{031a}
\end{equation}%
where $\mu _{i}\left( \mathbf{x}\right) =\sum_{j=1}^{4}u_{\left( j\right)
,i}\Gamma _{j}\left( \mathbf{x}\right) $ is the $i$th component of the drift
vector $\mathbf{\mu }\left( \mathbf{x}\right) $ given by Eq. (\ref{014}), $%
A_{ij}\left( \mathbf{x}\right) =u_{\left( j\right) ,i}\sqrt{\Gamma
_{j}\left( \mathbf{x}\right) }$ is the $ij$th element of the matrix $\mathbf{%
A}$ given as
\begin{equation}
\mathbf{A}\left( \mathbf{x}\right) =\left(
\begin{array}{cccc}
1 & -\sqrt{x_{1}} & -\sqrt{bx_{1}} & \sqrt{cx_{1}^{2}x_{2}} \\
0 & 0 & \sqrt{bx_{1}} & -\sqrt{cx_{1}^{2}x_{2}}%
\end{array}%
\right) ,  \label{032a}
\end{equation}%
and $\left\{ \zeta _{j}\left( t\right) ,j=1,\cdot \cdot \cdot ,4\right\} $,
defined as $\zeta _{j}\left( t\right) =\lim_{dt\rightarrow 0}\mathit{N}%
\left( 0,1/dt\right) $, are independent white noises with zero-means and $%
\left\langle \zeta _{i}\left( t\right) \zeta _{k}\left( s\right)
\right\rangle =\delta _{ik}\delta \left( t-s\right) $. Moreover, explicit
calculation yields
\begin{equation}
\mathbf{A}\left( \mathbf{x}\right) \cdot \mathbf{A}^{\tau }\left( \mathbf{x}%
\right) =2N\cdot \mathbf{D}\left( \mathbf{x}\right)  \label{033a}
\end{equation}%
with $\mathbf{D}\left( \mathbf{x}\right) $ given by Eq. (\ref{015}). Thus,
the Langevin equation, Eq. (\ref{030a}), is equivalent to the Fokker-Planck
equation given by Eq. (\ref{013}) \cite{gillespie2,gillespie3}. In the limit
$N\rightarrow \infty $, the fluctuation term of Eq. (\ref{031a})\ can be
neglected, and we obtain the mean-field equation, \
\begin{equation}
\frac{dx_{i}\left( t\right) }{dt}=\mu _{i}\left( \mathbf{x}\right) ,\text{ }%
i=1,2.  \label{016}
\end{equation}

\subsection{Fokker-Planck to Lagevin equation}

One can also construct Langevin equation directly from the Fokker-Planck
equation. Based on Eq. (\ref{013}) we have
\begin{equation}
\frac{dx_{i}\left( t\right) }{dt}=\mu _{i}\left( \mathbf{x}\right) +\frac{1}{%
\sqrt{N}}\sum_{k=1}^{2}B_{ik}\left( \mathbf{x}\right) \xi _{k}\left(
t\right) ,\text{ }i=1,2,  \label{017}
\end{equation}%
where the matrix $\mathbf{B}$ is defined as $\mathbf{B=}\sqrt{2N\mathbf{D}}$%
, and $\xi _{1}$ and $\xi _{2}$ are two independent white noises with
zero-means and $\left\langle \xi _{i}\left( t\right) \xi _{k}\left( s\right)
\right\rangle =\delta _{ik}\delta \left( t-s\right) $. By employing the
matrix $\mathbf{D}$ of Eq. (\ref{015}) we can obtain the explicit form of $%
\mathbf{B}$ as
\begin{equation}
\mathbf{B=}\left( \frac{1}{2\sqrt{z_{1}^{2}+4z_{2}^{2}}}\right) \left(
\begin{array}{cc}
z_{+}\theta _{+}^{1/2}-z_{-}\theta _{-}^{1/2} & -2z_{2}\left( \theta
_{+}^{1/2}-\theta _{-}^{1/2}\right) \\
-2z_{2}\left( \theta _{+}^{1/2}-\theta _{-}^{1/2}\right) & -\left(
z_{-}\theta _{+}^{1/2}-z_{+}\theta _{-}^{1/2}\right)%
\end{array}%
\right) ,  \label{018}
\end{equation}%
where we introduce the short notations, $z_{1}=1+x_{1}$, $%
z_{2}=bx_{1}+cx_{1}^{2}x_{2}$, $z_{\pm }=z_{1}\pm \sqrt{z_{1}^{2}+4z_{2}^{2}}
$, and $\theta _{\pm }=z_{2}+z_{\pm }/2$ are the eigenvalues of the matrix $%
2N\mathbf{D}$.

We notice that the matrix $\mathbf{B}$ of Eq. (\ref{018}) is constructed by
assuming that the matrices $\mathbf{B}$ and $\mathbf{D}$ are in the same
vector space, this leads to two independent fluctuations in Langevinian
approach. On the other hand, the matrix $\mathbf{A}$ of Eq. (\ref{032a}) has
dimension $2\times 4$, and there are four independent fluctuations
associated with four chemical reaction channels in the system. However, two
different Langevin equation correspond to the same Fokker-Planck equation,
Eq. (\ref{013}) \cite{gillespie2,gillespie3}. Consequently, one can expect
two different Langevin equations should give the same results for the
statistical measures of the intrinsic fluctuations of the system, and this
is demonstrated analytically for linearized Langevin equations shown in the
next section.

\section{ Linearized Langevin equations}

The mean-field equation of Eq. (\ref{016})\ takes the form,
\begin{eqnarray}
\frac{d}{dt}x_{1} &=&1-x_{1}-x_{1}\left( b-cx_{1}x_{2}\right)  \notag \\
\frac{d}{dt}x_{2} &=&x_{1}\left( b-cx_{1}x_{2}\right) ,  \label{032}
\end{eqnarray}%
for which, the fixed point is $\mathbf{x}^{\ast ^{\tau }}=\left( x_{1}^{\ast
},x_{2}^{\ast }\right) =\left( 1,b/c\right) $. The stability of a fixed
point can be analyzed by the property of the eigenvalues associated with the
Jacobian matrix at the fixed point,
\begin{equation}
\mathbf{J}=\left(
\begin{array}{cc}
b-1 & c \\
-b & -c%
\end{array}%
\right) .  \label{033}
\end{equation}%
The eigenvalues may be complex conjugate to each other and denoted as $%
\lambda _{\pm }=\lambda _{R}\pm i\lambda _{I}$ with real $\lambda
_{R}=\left( b-1-c\right) /2$ and $\lambda _{I}=\sqrt{4c-\left( b-1-c\right)
^{2}}/2$. For $\lambda _{R}<0$ and $\lambda _{I}\neq 0$, the fixed point is
stable and the system moves spirally towards the fixed point in the time
course; on the other hand, the fixed point is unstable and the system moves
spirally away from the fixed point for $\lambda _{R}>0$ and $\lambda
_{I}\neq 0$. For the latter, when the system is away from the fixed point,
the trajectories may converge to a limit cycle. Then, the two cases are
separated by the line $\overline{\lambda }_{R}=0$ in the parametric space,
and the separation is referred as the Hopf bifurcation.

In the followings, we apply the linear response theory to the Langevin
equations, Eqs. (\ref{031a}) and (\ref{017}), and analyze the variations of
the distributions of molecular concentrations and the power spectra for the
spirally stable equilibrium states as the parameters change toward the Hopf
bifurcation line. Moreover, the results obtained from two Langevin
approaches are shown to be identical.

\subsection{Four-component white noise}

We linearize Eq. (\ref{031a}) about the equilibrium state $\mathbf{x}^{\ast
} $ for which, the parameters $b$ and $c$ have negative $\lambda _{R}$ and
real positive $\lambda _{I}$, and the result is
\begin{equation}
\frac{d}{dt}\mathbf{y}^{\left( 4\right) }\left( t\right) =\mathbf{J}\cdot
\mathbf{y}^{\left( 4\right) }\left( t\right) +\frac{1}{\sqrt{N}}\mathbf{A}%
\left( \mathbf{x}^{\ast }\right) \cdot \mathbf{\zeta }\left( t\right) ,
\label{034a}
\end{equation}%
where $\mathbf{y}^{\left( 4\right) \tau }=\left( x_{1}-x_{1}^{\ast
},x_{2}-x_{2}^{\ast }\right) $ with the superscript, $\left( 4\right) $,
denoting the case of two-component white noise, $\mathbf{J}$ is given by Eq.
(\ref{033}), $\mathbf{A}\left( \mathbf{x}^{\ast }\right) $ is given by Eq. (%
\ref{032a}) evaluated at the fixed point $\mathbf{x}^{\ast }$, and $\mathbf{%
\zeta }\left( t\right) $ is the four-component white noise with $\mathbf{%
\zeta }^{\tau }\left( t\right) =\left( \zeta _{1}\left( t\right) ,\cdot
\cdot \cdot ,\zeta _{4}\left( t\right) \right) $. The integral expression
for the solution of Eq. (\ref{034a}) becomes
\begin{equation}
\mathbf{y}^{\left( 4\right) }\left( t\right) =\frac{1}{\sqrt{N}}\int_{0}^{t}%
\left[ \exp \left( t-u\right) \mathbf{J}\right] \cdot \mathbf{A}\left(
\mathbf{x}^{\ast }\right) \cdot d\mathbf{W}^{\left( 4\right) }\left( u\right)
\label{035a}
\end{equation}%
in the frame of Ito calculus \cite{gardiner}, where the Wiener process $%
\mathbf{W}^{\left( 4\right) }\left( u\right) $, $\mathbf{W}^{\left( 4\right)
\tau }\left( u\right) =\left( w_{1}^{\left( 4\right) }\left( u\right) ,\cdot
\cdot \cdot ,w_{4}^{\left( 4\right) }\left( u\right) \right) $, is related
to the white noise $\mathbf{\zeta }\left( u\right) $ by $d\mathbf{W}^{\left(
4\right) }\left( u\right) /du=\mathbf{\zeta }\left( u\right) $, and the
initial conditions are set as $\mathbf{y}^{\left( 4\right) }\left( 0\right)
=0$ and $\mathbf{\zeta }\left( 0\right) =0$, that is, the system is at a
stable fixed point without fluctuations at the time $t=0$. \

We diagonalize the matrix $\mathbf{J}$ of Eq. (\ref{035a}) via the
transformation matrix $\mathbf{M}$,

\begin{equation}
\mathbf{J}=\mathbf{M}\cdot \left(
\begin{array}{cc}
-\lambda _{1} & 0 \\
0 & -\lambda _{2}%
\end{array}%
\right) \cdot \mathbf{M}^{-1}  \label{036a}
\end{equation}%
with $\lambda _{1}=-\left( \lambda _{R}-i\lambda _{I}\right) $ and $\lambda
_{2}=-\left( \lambda _{R}+i\lambda _{I}\right) $, where the entries of
matrix $\mathbf{M}$, $m_{ij}$ for $i,j=1,2$, are normalized to satisfy the
relation $m_{11}m_{22}-m_{12}m_{21}=1$. Then, the expressions for the
components $y_{1}^{\left( 4\right) }\left( t\right) $ and $y_{2}^{\left(
4\right) }\left( t\right) $ can be obtained by substituting the matrix $%
\mathbf{J}$ of Eq. (\ref{035a}) with the result of Eq. (\ref{038a}). By
introducing the Ito integral $I_{i}^{\left( 4\right) }\left( \gamma
,t\right) $ for the Wiener process $w_{i}^{\left( 4\right) }\left( t\right) $
as
\begin{equation}
I_{i}^{\left( 4\right) }\left( \gamma ,t\right) =\exp \left( -\gamma
t\right) \int_{0}^{t}\exp \left( \gamma u\right) dw_{i}^{\left( 4\right)
}\left( u\right)  \label{037a}
\end{equation}%
for $i=1,\cdot \cdot \cdot ,4$, we have
\begin{equation}
y_{1}^{\left( 4\right) }\left( t\right) =\frac{1}{\sqrt{N}}\sum_{k=1}^{4}%
\left[ F_{1k}\left( \mathbf{x}^{\ast }\right) I_{k}^{\left( 4\right) }\left(
\lambda _{1},t\right) +F_{2k}\left( \mathbf{x}^{\ast }\right) I_{k}^{\left(
4\right) }\left( \lambda _{2},t\right) \right] ,  \label{038a}
\end{equation}%
where we introduce the functions,
\begin{equation}
F_{1k}\left( \mathbf{x}^{\ast }\right) =\alpha _{1}A_{1k}\left( \mathbf{x}%
^{\ast }\right) +\overline{\alpha }_{1}A_{2k}\left( \mathbf{x}^{\ast
}\right) ,  \label{039a}
\end{equation}%
and%
\begin{equation}
F_{2k}\left( \mathbf{x}^{\ast }\right) =\alpha _{2}A_{1k}\left( \mathbf{x}%
^{\ast }\right) +\overline{\alpha }_{2}A_{2k}\left( \mathbf{x}^{\ast
}\right) .  \label{039b}
\end{equation}%
Here, $A_{jk}$ is the $\left( j,k\right) $th element of the matrix $\mathbf{A%
}$ of Eq. (\ref{032a}) for $j=1,2$ and $k=1,\cdot \cdot \cdot ,4$, and the
parameters $\alpha _{i}$ and $\overline{\alpha }_{i}$ are defined as $\alpha
_{1}=m_{11}m_{22}$, $\alpha _{2}=-m_{12}m_{21}$, $\overline{\alpha }%
_{1}=-m_{11}m_{12}$, and $\overline{\alpha }_{2}=m_{12}m_{11}$ with $m_{ij}$
given by the matrix $\mathbf{M}$ of Eq. (\ref{036a}).

The expression of Eq. (\ref{038a}) indicates that $y_{1}^{\left( 4\right)
}\left( t\right) $ is linearly proportional to Wiener processes, and this
leads to the vanishing mean values of $y_{1}^{\left( 4\right) }\left(
t\right) $, $\left\langle y_{1}^{\left( 4\right) }\left( t\right)
\right\rangle =0$. Then, we compute the variance of steady-state
distribution defined as
\begin{equation}
\sigma _{s}^{2}\left( y_{i}^{\left( 4\right) }\right) =\lim_{t\rightarrow
\infty }\left\{ \left\langle y_{i}^{\left( 4\right) 2}\left( t\right)
\right\rangle -\left\langle y_{1}^{\left( 4\right) }\left( t\right)
\right\rangle ^{2}\right\} ,\text{ }i=1,2,  \label{039c}
\end{equation}%
and the result is
\begin{equation}
\sigma _{s}^{2}\left( y_{1}^{\left( 4\right) }\right) =\frac{-1}{2N}%
\sum_{k=1}^{4}\left( \frac{F_{1k}^{2}\left( \mathbf{x}^{\ast }\right) }{%
\lambda _{R}-i\lambda _{I}}+\frac{F_{2k}^{2}\left( \mathbf{x}^{\ast }\right)
}{\lambda _{R}+i\lambda _{I}}+\frac{2F_{1k}\left( \mathbf{x}^{\ast }\right)
F_{2k}\left( \mathbf{x}^{\ast }\right) }{\lambda _{R}}\right) .  \label{040a}
\end{equation}%
By substituting the explicit forms of the functions $F_{ij}\left( \mathbf{x}%
^{\ast }\right) $ into Eq. (\ref{040a}), we have
\begin{equation}
\sigma _{s}^{2}\left( y_{1}^{\left( 4\right) }\right) =\frac{-1}{2N}\left[
\frac{\Psi _{1}\left( \mathbf{x}^{\ast }\right) }{\lambda _{R}-i\lambda _{I}}%
+\frac{\Psi _{2}\left( \mathbf{x}^{\ast }\right) }{\lambda _{R}+i\lambda _{I}%
}+\frac{\Psi _{3}\left( \mathbf{x}^{\ast }\right) }{\lambda _{R}}\right]
\label{040b}
\end{equation}%
with
\begin{equation}
\Psi _{1}\left( \mathbf{x}^{\ast }\right) =\left( 2+2b\right) \alpha
_{1}^{2}-4b\alpha _{1}\overline{\alpha }_{1}+2b\overline{\alpha }_{1}^{2}
\label{040c}
\end{equation}%
\begin{equation}
\Psi _{2}\left( \mathbf{x}^{\ast }\right) =\left( 2+2b\right) \alpha
_{1}^{2}-4b\alpha _{2}\overline{\alpha }_{2}+2b\overline{\alpha }_{2}^{2}
\label{040d}
\end{equation}%
and
\begin{equation}
\Psi _{3}\left( \mathbf{x}^{\ast }\right) =\left( 4+4b\right) \alpha
_{1}\alpha _{2}-4b\left( \alpha _{2}\overline{\alpha }_{1}+\alpha _{1}%
\overline{\alpha }_{2}-\overline{\alpha }_{1}\overline{\alpha }_{2}\right) .
\label{040e}
\end{equation}%
Note that the result of $\sigma _{s}^{2}\left( y_{2}^{\left( 4\right)
}\right) $ takes the same form as $\sigma _{s}^{2}\left( y_{1}^{\left(
4\right) }\right) $ but with the replacement, $\alpha _{i}$ $\rightarrow $ $%
\beta _{i}$ and $\overline{\alpha }_{i}$ $\rightarrow $ $\overline{\beta }%
_{i}$. We also notice that Eq. (\ref{040b}) clearly indicates the existence
of a pole at the Hopf bifurcation line $\lambda _{R}=0$ for the variance of
steady-state distribution.

Additional kinematic features caused by intrinsic fluctuations can be
revealed from the power spectra of dynamical variables. By taking the
Fourier transform of $y_{i}^{\left( 4\right) }\left( t\right) $,
\begin{equation}
\overline{y}_{i}^{\left( 4\right) }\left( \omega ,T\right)
=\int_{0}^{T}dt\exp \left( -i\omega t\right) y_{i}^{\left( 4\right) }\left(
t\right) ,  \label{041a}
\end{equation}%
we define the spectrum as
\begin{equation}
S_{i}^{\left( 4\right) }\left( \omega \right) =\lim_{T\rightarrow \infty }%
\frac{1}{2\pi T}\left\langle \left\vert \overline{y}_{i}^{\left( 4\right)
}\left( \omega ,T\right) \right\vert ^{2}\right\rangle  \label{042a}
\end{equation}%
for $i=1$ and $2$, where the average is taken over the Wiener processes.

The typical terms in $\overline{y}_{i}^{\left( 4\right) }\left( \omega
,T\right) $ are the Fourier transforms of Ito integrals, \ \
\begin{equation}
\int_{0}^{T}I_{i}^{\left( 4\right) }\left( \gamma ,t\right) \exp \left(
-i\omega t\right) dt=\left( \frac{1}{\gamma +i\omega }\right) \overline{I}%
_{i}^{\left( 4\right) }\left( \omega ,T\right)  \label{043a}
\end{equation}%
with
\begin{equation}
\overline{I}_{i}^{\left( 4\right) }\left( \omega ,T\right) =\int_{0}^{T}\exp %
\left[ -i\omega s\right] dw_{i}^{\left( 4\right) }\left( s\right)
\label{044a}
\end{equation}%
for sufficiently large $T$. Then, based on Eq. (\ref{038}) we have \
\begin{equation}
\overline{y}_{1}^{\left( 4\right) }\left( \omega ,T\right) =\frac{-1}{\sqrt{N%
}}\sum_{k=1}^{4}\left( \frac{F_{1k}\left( \mathbf{x}^{\ast }\right) }{%
\lambda _{R}-i\left( \omega +\lambda _{I}\right) }+\frac{F_{2k}\left(
\mathbf{x}^{\ast }\right) }{\lambda _{R}-i\left( \omega -\lambda _{I}\right)
}\right) \overline{I}_{k}^{\left( 4\right) }\left( \omega ,T\right)
\label{045a}
\end{equation}%
for sufficiently large $T$. By using Eq. (\ref{045a}) and the equality
\begin{equation}
\lim_{T\rightarrow \infty }\frac{1}{2\pi T}\left\langle \overline{I}%
_{i}^{\left( 4\right) }\left( \omega ,T\right) \overline{I}_{k}^{\left(
4\right) \ast }\left( \omega ,T\right) \right\rangle =\frac{1}{2\pi }\delta
_{ik}  \label{046a}
\end{equation}%
for Eq. (\ref{042a}), we obtain
\begin{equation}
S_{1}^{\left( 4\right) }\left( \omega \right) =\left( \frac{1}{2\pi N}%
\right) \sum_{k=1}^{4}\left\vert \frac{F_{1k}\left( \mathbf{x}^{\ast
}\right) }{\lambda _{R}-i\left( \omega +\lambda _{I}\right) }+\frac{%
F_{2k}\left( \mathbf{x}^{\ast }\right) }{\lambda _{R}-i\left( \omega
-\lambda _{I}\right) }\right\vert ^{2},  \label{047a}
\end{equation}%
where $\delta _{ik}$ is the Kronecker delta of $i$ and $k$. By working out
the form of Eq. (\ref{047a}) algebraically, we have
\begin{eqnarray}
S_{1}^{\left( 4\right) }\left( \omega \right) &=&\left( \frac{1}{2\pi N}%
\right) \left\{ \frac{\Phi _{1}\left( \mathbf{x}^{\ast }\right) }{\lambda
_{R}^{2}+\left( \omega +\lambda _{I}\right) ^{2}}+\frac{\Phi _{2}\left(
\mathbf{x}^{\ast }\right) }{\omega ^{2}-\left( \lambda _{I}+i\lambda
_{R}\right) ^{2}}\right.  \notag \\
&&\left. +\frac{\Phi _{3}\left( \mathbf{x}^{\ast }\right) }{\omega
^{2}-\left( \lambda _{I}-i\lambda _{R}\right) ^{2}}+\frac{\Phi _{4}\left(
\mathbf{x}^{\ast }\right) }{\lambda _{R}^{2}+\left( \omega -\lambda
_{I}\right) ^{2}}\right\}  \label{048a}
\end{eqnarray}%
with
\begin{equation}
\Phi _{1}\left( \mathbf{x}^{\ast }\right) =\left( 2+2b\right) \alpha
_{1}\alpha _{1}^{\ast }-2b\left( \alpha _{1}\overline{\alpha }_{1}^{\ast }+%
\overline{\alpha }_{1}\alpha _{1}^{\ast }-\alpha _{1}^{\ast }\overline{%
\alpha }_{1}^{\ast }\right)  \label{048b}
\end{equation}%
\begin{equation}
\Phi _{2}\left( \mathbf{x}^{\ast }\right) =\left( 2+2b\right) \alpha
_{1}\alpha _{2}^{\ast }-2b\left( \overline{\alpha }_{1}\alpha _{2}^{\ast
}+\alpha _{1}\overline{\alpha }_{2}^{\ast }-\overline{\alpha }_{1}\overline{%
\alpha }_{2}^{\ast }\right)  \label{048c}
\end{equation}%
\begin{equation}
\Phi _{3}\left( \mathbf{x}^{\ast }\right) =\left( 2+2b\right) \alpha
_{1}^{\ast }\alpha _{2}-2b\left( \alpha _{1}^{\ast }\overline{\alpha }_{2}+%
\overline{\alpha }_{1}^{\ast }\alpha _{2}-\overline{\alpha }_{1}^{\ast }%
\overline{\alpha }_{2}\right)  \label{048d}
\end{equation}%
and
\begin{equation}
\Phi _{4}\left( \mathbf{x}^{\ast }\right) =\left( 2+2b\right) \alpha
_{2}\alpha _{2}^{\%}-2b\left( \alpha _{2}^{\ast }\overline{\alpha }%
_{2}+\alpha _{2}\overline{\alpha }_{2}^{\ast }-\overline{\alpha }_{2}%
\overline{\alpha }_{2}^{\ast }\right) .  \label{048e}
\end{equation}%
Similarly, $S_{2}^{\left( 4\right) }\left( \omega \right) $ can be obtained
from the results of $S_{1}^{\left( 4\right) }\left( \omega \right) $ by the
replacements $\alpha _{i}\rightarrow \beta _{i}$ and $\overline{\alpha }%
_{i}\rightarrow \overline{\beta }$. The result of Eq. (\ref{048a}) indicates
that the power spectra develop a pole at $\omega =\lambda _{I}$ as the
parameters are tuned toward the Hopf bifurcation line $\lambda _{R}=0$. \

\subsection{Two-component white noise\ }

The Langevin equation of Eq. (\ref{017}) is linearized about the equilibrium
state $\mathbf{x}^{\ast }$ to yield

\begin{equation}
\frac{d}{dt}\mathbf{y}^{\left( 2\right) }\left( t\right) =\mathbf{J}\cdot
\mathbf{y}^{\left( 2\right) }\left( t\right) +\frac{1}{\sqrt{N}}\mathbf{B}%
\left( \mathbf{x}^{\ast }\right) \cdot \mathbf{\xi }\left( t\right) ,
\label{036}
\end{equation}%
where $\mathbf{y}^{\left( 2\right) \tau }=\left( x_{1}-x_{1}^{\ast
},x_{2}-x_{2}^{\ast }\right) $ with the superscript, $\left( 2\right) $,
denoting the case of two-component white noise, corresponds to the vector $%
\mathbf{y}^{\left( 4\right) }$ of Eq. (\ref{034a}), and $\mathbf{B}\left(
\mathbf{x}^{\ast }\right) $ is given by Eq. (\ref{018}) evaluated at the
fixed point $\mathbf{x}^{\ast }$. We first express the solution of Eq. (\ref%
{036}) as

\begin{equation}
\mathbf{y}^{\left( 2\right) }\left( t\right) =\frac{1}{\sqrt{N}}\int_{0}^{t}%
\left[ \exp \left( t-u\right) \mathbf{J}\right] \cdot \mathbf{B}\left(
\mathbf{x}^{\ast }\right) \cdot d\mathbf{W}^{\left( 2\right) }\left( u\right)
\label{037}
\end{equation}%
\ in the frame of Ito calculus \cite{gardiner}, where the Wiener process $%
\mathbf{W}^{\left( 2\right) }\left( u\right) $, $\mathbf{W}^{\left( 2\right)
\tau }\left( u\right) =\left( w_{1}^{\left( 2\right) }\left( u\right)
,w_{2}^{\left( 2\right) }\left( u\right) \right) $, is related to the white
noise $\mathbf{\xi }\left( u\right) $ by $d\mathbf{W}^{\left( 2\right)
}\left( u\right) /du=\mathbf{\xi }\left( u\right) $, and the initial
conditions are set as $\mathbf{y}^{\left( 2\right) }\left( 0\right) =0$ and $%
\mathbf{\xi }\left( 0\right) =0$. Then, by following the same process for
the case of four-component white noise, we can obtain the first two moments
of the steady-state probability density distribution in a straightforward
way. The mean values vanish, $\left\langle y_{i}^{\left( 2\right) }\left(
t\right) \right\rangle =0$ for $i=1,2$, and the variance for $y_{1}^{\left(
2\right) }$ is
\begin{equation}
\sigma _{s}^{2}\left( y_{1}^{\left( 2\right) }\right) =\frac{-1}{2N}%
\sum_{k=1}^{2}\left( \frac{G_{k1}^{2}\left( \mathbf{x}^{\ast }\right) }{%
\lambda _{R}-i\lambda _{I}}+\frac{G_{k2}^{2}\left( \mathbf{x}^{\ast }\right)
}{\lambda _{R}+i\lambda _{I}}+\frac{2G_{k1}\left( \mathbf{x}^{\ast }\right)
G_{k2}\left( \mathbf{x}^{\ast }\right) }{\lambda _{R}}\right) ,  \label{040}
\end{equation}%
where the functions $G_{ij}\left( \mathbf{x}^{\ast }\right) $ for $i,j=1,2$
are
\begin{equation}
G_{i1}\left( \mathbf{x}^{\ast }\right) =\alpha _{1}B_{1i}\left( \mathbf{x}%
^{\ast }\right) +\overline{\alpha }_{1}B_{2i}\left( \mathbf{x}^{\ast }\right)
\label{038-1}
\end{equation}%
and%
\begin{equation}
G_{i2}\left( \mathbf{x}^{\ast }\right) =\alpha _{2}B_{1i}\left( \mathbf{x}%
^{\ast }\right) +\overline{\alpha }_{2}B_{2i}\left( \mathbf{x}^{\ast }\right)
\label{038-2}
\end{equation}%
with $B_{ij}\left( \mathbf{x}^{\ast }\right) $ the $ij$th element of the
matrix $\mathbf{B}$ of Eq. (\ref{018}) evaluated at the fixed point $\mathbf{%
x}^{\ast }$. Note that $\sigma _{s}^{2}\left( y_{2}^{\left( 2\right)
}\right) $ also possesses the same form as $\sigma _{s}^{2}\left(
y_{1}^{\left( 2\right) }\right) $ but with the replacements $\alpha
_{i}\rightarrow \beta _{i}$ and $\overline{\alpha }_{i}\rightarrow \overline{%
\beta }$. By substituting the explicit forms of the functions $G_{ij}\left(
\mathbf{x}^{\ast }\right) $ into Eq. (\ref{040}), our algebraic results give
the identity, $\sigma _{s}^{2}\left( y_{1}^{\left( 2\right) }\right) =\sigma
_{s}^{2}\left( y_{1}^{\left( 4\right) }\right) $ with $\sigma _{s}^{2}\left(
y_{1}^{\left( 4\right) }\right) $ given by Eq. (\ref{040b}). Consequently,
we also have $\sigma _{s}^{2}\left( y_{2}^{\left( 2\right) }\right) =\sigma
_{s}^{2}\left( y_{2}^{\left( 4\right) }\right) $.

The power spectra can also be calculated by following the same process as
the case of four-component white noise. The Fourier transform of $%
y_{1}^{\left( 2\right) }\left( t\right) $ is
\begin{equation}
\overline{y}_{1}^{\left( 2\right) }\left( \omega ,T\right)
=\int_{0}^{T}dt\exp \left( -i\omega t\right) y_{1}^{\left( 2\right) }\left(
t\right) ,  \label{041}
\end{equation}%
and the spectrum is defined as
\begin{equation}
S_{1}^{\left( 2\right) }\left( \omega \right) =\lim_{T\rightarrow \infty }%
\frac{1}{2\pi T}\left\langle \left\vert \overline{y}_{1}^{\left( 2\right)
}\left( \omega ,T\right) \right\vert ^{2}\right\rangle .  \label{042}
\end{equation}%
The expression of Eq. (\ref{037}) can be used to obtain
\begin{equation*}
\overline{y}_{1}^{\left( 2\right) }\left( \omega ,T\right) =\frac{-1}{\sqrt{N%
}}\sum_{k=1}^{2}\left( \frac{G_{k1}\left( \mathbf{x}^{\ast }\right) }{%
\lambda _{R}-i\left( \omega +\lambda _{I}\right) }+\frac{G_{k2}\left(
\mathbf{x}^{\ast }\right) }{\lambda _{R}-i\left( \omega -\lambda _{I}\right)
}\right) \overline{I}_{k}^{\left( 2\right) }\left( \omega ,T\right)
\end{equation*}%
with
\begin{equation}
\overline{I}_{k}^{\left( 2\right) }\left( \omega ,T\right) =\int_{0}^{T}\exp %
\left[ -i\omega s\right] dw_{k}^{\left( 2\right) }\left( s\right)
\label{043}
\end{equation}%
for sufficiently large $T$. Then, we have
\begin{equation}
S_{1}^{\left( 2\right) }\left( \omega \right) =\left( \frac{1}{2\pi N}%
\right) \sum_{k=1}^{2}\left\vert \frac{G_{k1}\left( \mathbf{x}^{\ast
}\right) }{\lambda _{R}-i\left( \omega +\lambda _{I}\right) }+\frac{%
G_{k2}\left( \mathbf{x}^{\ast }\right) }{\lambda _{R}-i\left( \omega
-\lambda _{I}\right) }\right\vert ^{2}  \label{046}
\end{equation}%
for the power spectrum of Eq. (\ref{042}). Explicit algebraic computations
for Eq. (\ref{046}) yield the result $S_{1}^{\left( 2\right) }\left( \omega
\right) =S_{1}^{\left( 4\right) }\left( \omega \right) $ with $S_{1}^{\left(
4\right) }\left( \omega \right) $ given by Eq. (\ref{048a}). Similarly, one
can also expect that the power spectrum for the dynamic variable $%
y_{2}^{\left( 2\right) }\left( t\right) $, $S_{2}^{\left( 2\right) }\left(
\omega \right) $, has the equality $S_{2}^{\left( 2\right) }\left( \omega
\right) =S_{2}^{\left( 4\right) }\left( \omega \right) $ with $S_{2}^{\left(
4\right) }\left( \omega \right) $ obtained from the calculations of
four-component noise. \

\section{Numerical Results}

Numerical calculations, based on different frameworks, are carried out for
two quantities, probability density distributions and power spectra.
Different frameworks may yield distinguished results, and we focus on the
differences caused by the molecular number and the distance away from the
Hopf bifurcation line $\lambda _{R}=0$ for a stable equilibrium state in
deterministic dynamics. As two forms of Langevin equations are shown to be
equivalent, we take Eq. (\ref{034a}) for the Langevin approach. Firstly, a
variety of molecular trajectories with the same initial condition, $%
x_{1}\left( 0\right) =x_{1}^{\ast }$ and $x_{2}\left( 0\right) =x_{2}^{\ast
} $, are generated from the master and the Langevin equation. The master
equation, given by Eq. (\ref{007}), is the primitive approach and provides
the description of the system at the level of individual molecules, and we
use the Gillespie algorithm to generate trajectories \cite{gillespie4},
meanwhile the molecular trajectories of Eq. (\ref{034a}) are generated by
using the standard simulation technique for independent Gaussian random
numbers. Then, we construct the histograms of different states by sampling
the data of trajectories and obtain the steady-state probability density
distributions, and the Fourier transforms of the trajectories are computed
to obtain the power spectra. There are two parameters, $b$ and $c$, for
numerical calculations, we fix the parameter $c=1$ and vary the $b$ value to
have different $\lambda _{R}$ values, $\lambda _{R}=-1+b/2$.

The steady-state probability density distributions, obtained from master
equation $P_{s}^{M}\left( y_{1}\right) $ and from Langevin equation, $%
P_{s}^{L}\left( y_{1}\right) $ as functions of $y_{1}$ are shown in Fig. 1.
Here, the distributions all are normalized to $1$,
\begin{equation}
\int_{-\infty }^{\infty }P_{s}\left( y_{1}\right) dy_{1}=1.  \label{047aa}
\end{equation}%
To give a quantitative measure about the difference between $P_{s}^{L}\left(
y_{1}\right) $ and $P_{s}^{M}\left( y_{1}\right) $, we introduce the
deviation $\Delta _{N}\left( \lambda _{R}\right) $ defined as
\begin{equation}
\Delta _{N}\left( \lambda _{R}\right) =\int_{-\infty }^{\infty }\left\vert
P_{s}^{L}\left( y_{1}\right) -P_{s}^{M}\left( y_{1}\right) \right\vert dy_{1}
\label{047}
\end{equation}%
for given values of $\lambda _{R}$ and $N$. Based on the distributions shown
in Fig.1, we have $\Delta _{200}\left( -0.5\right) =0.0585$, $\Delta
_{600}\left( -0.5\right) =0.0347$, and$\ \Delta _{1200}\left( -0.5\right)
=0.0251$ for Fig. 1(a), $\Delta _{200}\left( -0.1\right) =0.1834$, $\Delta
_{600}\left( -0.1\right) =0.1197$, and $\Delta _{1200}\left( -0.1\right)
=0.0894$ for Fig. 1(b), and $\Delta _{200}\left( -0.01\right) =0.2445$, $%
\Delta _{600}\left( -0.01\right) =0.1823$, and $\Delta _{1200}\left(
-0.01\right) =0.1500$ for Fig. 1(c). In general, the deviation of $%
P_{s}^{L}\left( y_{1}\right) $ from $P_{s}^{M}\left( y_{1}\right) $ is
expected to be noticeable for small system size $N$. However, our results
indicate that the difference between $P_{s}^{L}\left( y_{1}\right) $ and $%
P_{s}^{M}\left( y_{1}\right) $ is small for systems far away from the Hopf
bifurcation line $\lambda _{R}=0$ even with small $N$. For example, the $%
\Delta _{N}\left( \lambda _{R}\right) $ value with $\lambda _{R}=-0.5$ is
less than $6$ percentage of the distribution when the system size is reduced
down to $N=200$, namely, $\Delta _{200}\left( -0.5\right) =0.0585$. On the
other hand, noticeable difference between two distributions is observed for
systems closed to the bifurcation line $\lambda _{R}=0$ even with large $N$.
For example, the $\Delta _{N}\left( \lambda _{R}\right) $ value with $%
\lambda _{R}=-0.01$ still has $15$ percentage of the distribution when the
system size is increased up to $N=1200$, namely, $\Delta _{1200}\left(
-0.01\right) =0.1500$. Thus, the $\lambda _{R}$ value of equilibrium state
may play a more important role than the system size in determining which
formulation is adequate for the study of stochasticity in chemical reactions.

\begin{figure}
\includegraphics[width=0.8\textwidth]{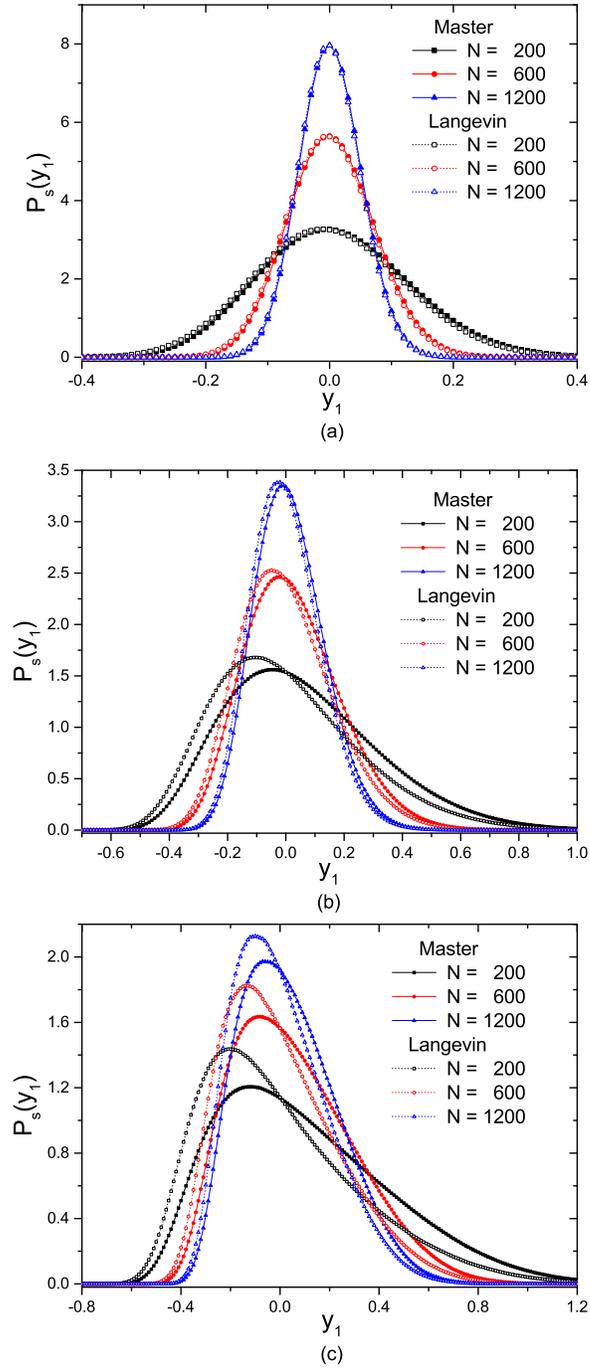}
\bigskip
\bigskip
\caption{The steady-state probability density distributions $P_{s}\left( y_{1}\right)
$, obtained from the master and the Langevin equations, as functions of $%
y_{1}$: $\left( a\right) $ $b=1.0$ and $\lambda _{R}=-0.5$, $\left( b\right)
$ $b=1.8$ and $\lambda _{R}=-0.1$, and $\left( c\right) $ $b=1.98$ and $%
\lambda _{R}=-0.01$ with $N=200$, $600$, and $1200$.}
\end{figure}

The variance of steady-state probability density distribution is calculated
and analyzed to reveal more informations about the distributions in
different formulations, in particular, about the reliability of linearized
Langevin equation. We show the logarithm of variance, $\ln \sigma
_{s}^{2}\left( y_{1}\right) $, as a function of the logarithm of $\lambda
_{R}$, $\ln \lambda _{R}$, for systems with $N=200$, $600$, and $1200$ in
Fig. 2. Our results indicate that the variances obtained from master
equation are, in general, larger than those obtained from Langevin equation.
Note that the variances obtained from linearized Langevin are even larger
than those from master, and there are big deviations from the results of
master and Langevin for $0>\lambda _{R}>-0.25$ with $N=200$, $0>\lambda
_{R}>-0.15$ with $N=600$, and $0>\lambda _{R}>-0.1$ with $N=1200$, and the
results tend to diverge for $\lambda _{R}\rightarrow 0_{-}$ as shown in Fig.
2. Thus, the linearization scheme of Langevin equation becomes highly
unreliable for systems with very small $\left\vert \lambda _{R}\right\vert $
and $\lambda _{R}<0$. \

\begin{figure}
\includegraphics[width=0.80\textwidth]{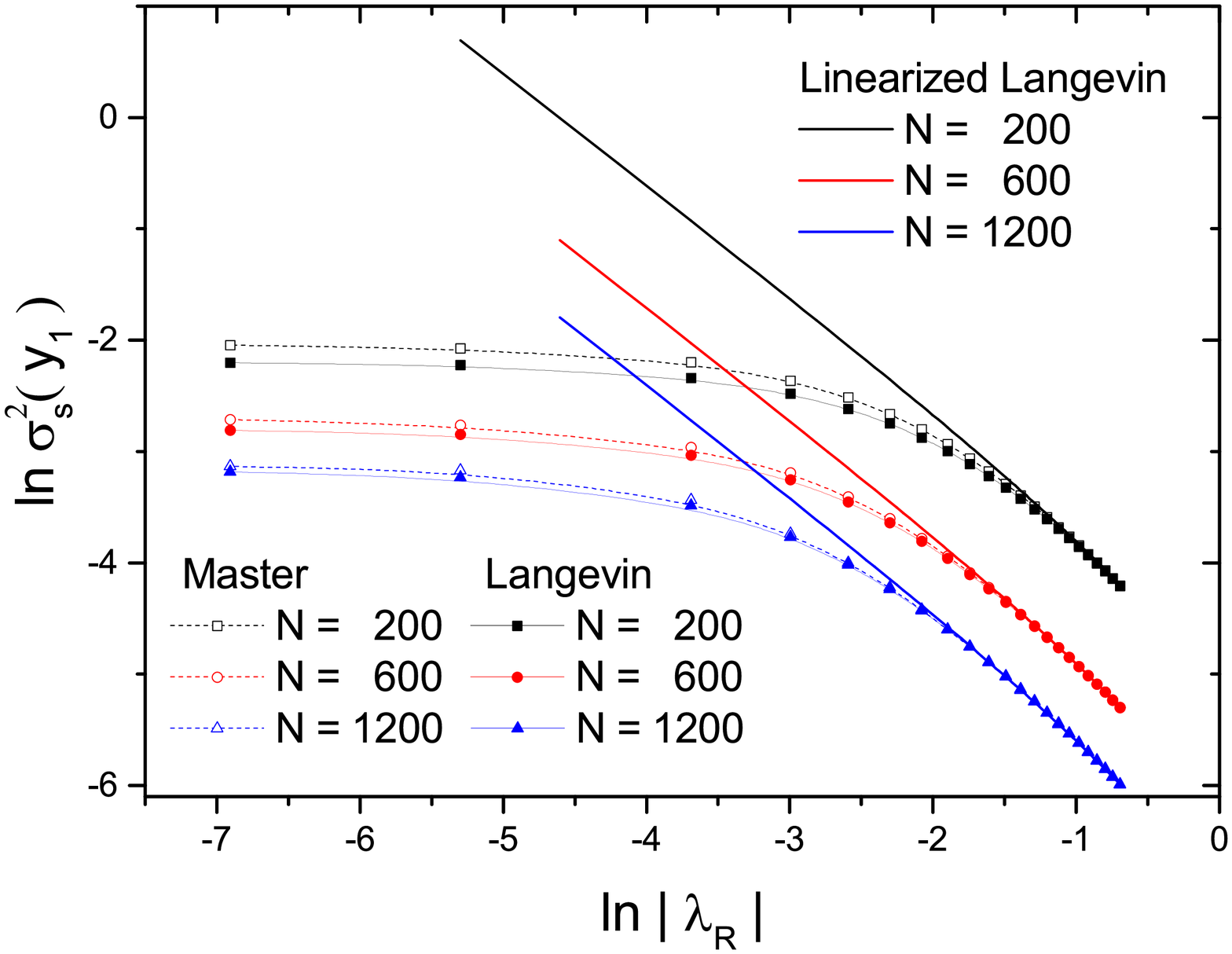}
\bigskip
\bigskip
\caption{The logarithm of variance of steady-state probability density distribution, $%
\ln \sigma _{s}^{2}\left( y_{1}\right) $, obtained from the master, the
Langevin, and the linearized Langevin equations, as a function of the
logarithm of the absolute value of $\lambda _{R}$, $\ln \left\vert \lambda
_{R}\right\vert $, for systems with $N=200$, $600$, and $1200$. The results
of linearized Langevin equation are obtained from Eq. (\ref{040b}).}
\end{figure}

The power spectrum provides another aspect for the kinematic properties of
systems. Since power spectrum defined in frequency space is complementary to
probability density distribution defined in state space, a steady-state
probability density distribution with smaller variance would correspond to
the power spectrum covering a wider range of frequency. The results of power
spectra for systems with $N=200$ and $1200$ are shown as ln$\left(
NS_{1}\left( \omega \right) \right) $ vs. $\omega $ for $\lambda _{R}=-0.5$,
$-0.1$, and $-0.01$ in Fig. 3. As shown in Figs. 3(c) and 3(d), the spectra
obtained from linearized Langevin equation deviate significantly from those
obtained from master and Langevin equations. Moreover, the peaks of the
spectra from three different formulations locate at $\omega =\lambda _{I}$
for systems closed to the bifurcation line $\lambda _{R}=0$, this is
consistent with the analytic result of linearized Langevin equation given by
Eq. (\ref{048a}) which clearly indicates that the spectrum develops a pole
of second order as $1/\left( \omega -\lambda _{I}\right) ^{2}$ at the
bifurcation line $\lambda _{R}=0$, although the pole is absent for master
and Langevin equations. \ \

\begin{figure}
\includegraphics[width=1.00\textwidth]{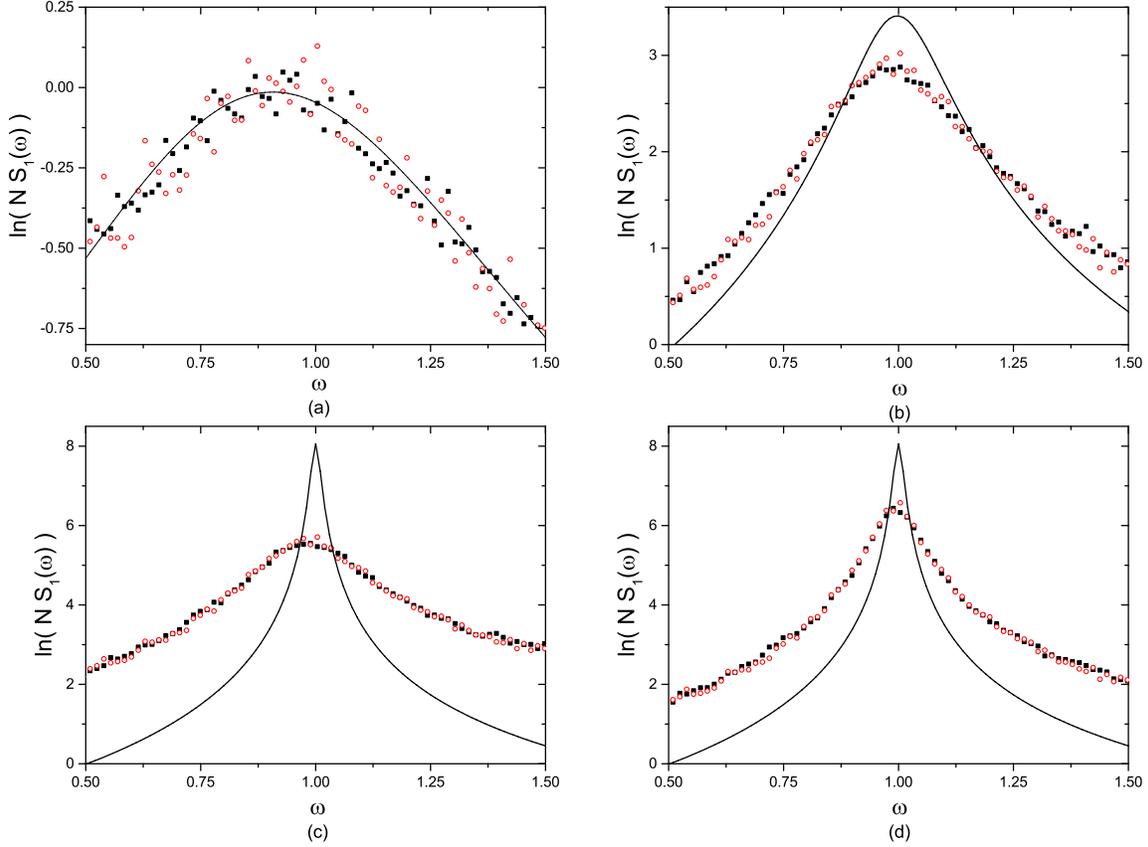}
\bigskip
\bigskip
\caption{The logarithm of power spectrum, $\ln \left( NS_{1}\left( \omega \right)
\right) $, obtained from the master (squares), the Langevin (circles), and
linearized Langevin (solid lines) equations, as a function of $\omega $: $%
\left( a\right) $ $\lambda _{R}=-0.5$ and $N=200$, $\left( b\right) $ $%
\lambda _{R}=-0.1$ and $N=200$, $\left( c\right) $ $\lambda _{R}=-0.01$ and $%
N=200$, and $\left( d\right) $ $\lambda _{R}=-0.01$ and $N=1200$. The
results of linearized Langevin equation are obtained from Eq. (\ref{048a}).}
\end{figure}

\label{s8}

\section{Summary \ }

\label{s5-1}

Three different formulations, including master, Langevin, and linearized
Langevin equations, are used to analyze the effect of intrinsic fluctuations
for chemical reactions defined by the Brusselator model. The systems are
assumed to be in the phase of spirally stable fixed point for the
deterministic mean-field equation, and we analyze the effect of intrinsic
fluctuations based on steady-state probability density distributions in
state space and power spectra in frequency space. Moreover, the differences
between the results obtained from three formulations are investigated by
considering two factors, the system size and the distance from the Hopf
bifurcation line $\lambda _{R}=0$ for a spirally stable equilibrium state.

Our results indicate that the effect of intrinsic fluctuations based on
master equation gives larger variance in steady-state probability density
distribution than that obtained from Langevin equation, and the difference
in the variance of distribution is enhanced when the system size is reduced.
Moreover, the difference between the results of two formulations increases
significantly when the equilibrium state is closed to the bifurcation line $%
\lambda _{R}=0$. In general, the discrepancy between the results of master
and Langevin equations caused by the different distances of equilibrium
states from the bifurcation line is more noticeable than that caused by the
different system sizes. \

Our results also show that the effect of intrinsic fluctuations revealed
from linearized Langevin equation agrees very well with those obtained from
Langevin equation for system far away from the bifurcation line. However,
the linearization scheme of Langevin equation becomes inadequate for system
closed to the bifurcation line. As $\lambda _{R}\rightarrow 0_{-}$, our
analytic results indicate that the variance associated with the steady-state
probability density function possesses a divergence as $1/\left\vert \lambda
_{R}\right\vert $ and the power spectrum tends to diverge at $\omega =$\ $%
\lambda _{I}$ as $1/\left( \omega -\lambda _{I}\right) ^{2}$; these singular
behaviors are absent in the results obtained from master and Langevin
equations.

In conclusion, our results provide insights on the adequacy of different
approaches for taking account of the intrinsic fluctuations into a system.
Although the study is based on the Brusselator model, our results about the
discrepancy between three frameworks can be quite general.

\label{f26}

\label{h-26}


\begin{thebibliography}{99}
\bibitem{crawford} J.D. Crawford, Rev. Mod. Phys., \textbf{63}, 991, (1991).

\bibitem{arnold} V.I. Arnold, \emph{Geometrical Methods in the Theory of
Ordinary Differential Equations}, Springer-Verlag New York(1983).

\bibitem{strogatz} S.H. Strogatz, \emph{Nonlinear Dynamics and Chaos}, 2nd
edn, Westview Press (2014).

\bibitem{tomita} K. Tomita, T. Ohta, and H. Tomita, Prog. Theor. Phys.,%
\textbf{\ 52}, 1744, (1974).

\bibitem{boland} R.P. Boland, T. Galla, and A.J. McKane, J. Stat. Mech.:
Theory and Experiment, \textbf{9}, P09001, (2008).

\bibitem{mckane} A.J. McKane and T.J. Newman, Phys. Rev. Lett., \textbf{94},
218102, (2005).

\bibitem{moreira} A.A. Moreira, A. Mathur, D. Diermeier, and L.A.N. Amaral,
Proc. Natl. Acad. Sci., \textbf{101}, 12085, (2004).

\bibitem{lama} M.S. de la Lama, I.G. Szendro, J.R. Iglesias, and H.S. Wio,
Eur. Phys. J., \textbf{B51}, 435, (2006).

\bibitem{scott} M. Scott, B. Ingalls, and M. Kaern, Chaos, \textbf{16},
026107, (2006).

\bibitem{moss} F. Moss and P.V.E. McClintock, \emph{Noise in Nonlinear
Dynamics}, Cambridge University Press, Cambridge(1989)

\bibitem{kampen} N.G. van Kampen, \emph{Stochastic Processes in Physics and
Chemistry}, 3rd edn, Elsevier, Amsterdam(2007).

\bibitem{sagues} F. Sagues, J.M. Sancho, and J. Garcia-Ojalvo, Rev. Mod.
Phys. \textbf{79}, 829, (2007).

\bibitem{gardiner} C.W. Gardiner, \emph{Handbook of Stochastic Method for
Physics, Chemistry and the Natural Sciences}, Spring-Verlag New York(1994).

\bibitem{risken} H. Risken, \emph{The Fokker-Planck Equation: Methods of
Solution and Applications}, Springer, Berlin (1989)

\bibitem{gaspard} P. Gaspard, J. Chem. Phys. \textbf{117}, 8905, (2002).

\bibitem{nakanishi} H. Nakanishi, T. Sakaue, and J. Wakou, J. Chem. Phys.
\textbf{139}, 214105, (2013).

\bibitem{broeck} C. Van den Broeck, M. Malek, and F. Baras, J. Stat. Phys.
\textbf{28}, 557, (1982).

\bibitem{vance} W. Vance and J. Ross, J. Chem. Phys. \textbf{105 }(2), 479,
(1996).

\bibitem{mori} F. Mori and A.S. Mikhailov, Phys. Rev. \textbf{E93}, 062206,
(2016).

\bibitem{xue} C. Xue and N. Goldenfeld, Phy. Rev. Lett., \textbf{119},
268101, (2017). \

\bibitem{gillespie1} D.T. Gillespie, J. Chem. Phys. \textbf{113}, 297,
(2000).

\bibitem{gillespie2} D.T. Gillespie, J. Phys. Chem. A \textbf{106}, 5063,
(2002).

\bibitem{gillespie3} D.T. Gillespie, Am. J. Phys. \textbf{64}, 1246, (1996).

\bibitem{gillespie4} D.T. Gillespie, Annu. Rev. Phys. Chem. \textbf{58}, 35,
(2007).
\end{thebibliography}
\end{document}